\documentclass[%
 reprint,
showpacs,
 amsmath,amssymb,
 aps,
 prl,
 floats,
]{revtex4-1}

\usepackage{graphicx}
\usepackage{dcolumn}
\usepackage{bm}
\usepackage{color}
\usepackage{soul}

\begin{document}


\title{Terahertz excitation of a coherent three-level $\Lambda$-type exciton-polariton microcavity mode}

\author{J.L. Tomaino$^1$}
\author{A.D. Jameson$^1$}
\author{{Yun-Shik} Lee$^1$}
\email{leeys@physics.oregonstate.edu}

\author{G. Khitrova$^2$}
\author{H.M. Gibbs$^2$}

\author{A.C. Klettke$^3$}
\author{M. Kira$^3$}
\author{S.W. Koch$^3$}
\affiliation{%
$^1$Department of Physics, Oregon State University, Corvallis,
Oregon 97331 \\
$^2$Optical Sciences Center, University of Arizona, Tucson,
Arizona 85721 \\
$^3$Department of Physics and Material Sciences Center,
Philipps-University, 35032 Marburg, Germany
}%


\date{\today}

\begin{abstract}
Interactions of few-cycle terahertz pulses with the induced optical polarization in a quantum-well microcavity reveal that the lower and higher exciton-polariton modes together with the optically forbidden $2p$-exciton state form a unique $\Lambda$-type three-level system.  Pronounced nonlinearities are observed via time-resolved strong-terahertz and weak-optical excitation spectroscopy and explained with a fully microscopic theory.
The results show that the terahertz pulses strongly couple the exciton-polariton states to the $2p$-exciton state while no resonant transition between the two polariton levels is observed.
\end{abstract}

\pacs{78.47.J-, 78.47.jb, 71.35.Cc, 42.55.Sa}

\maketitle


An optical microcavity enclosing semiconductor quantum-wells (QWs) is an elegant material system to harness light-matter interactions. When the exciton and cavity resonances are nearly resonant, they become strongly coupled giving rise to the so-called exciton-polariton modes~\cite{Khitrova:99}. Many fundamental questions concerning quantum optical phenomena in semiconductors---from cavity QED~\cite{Lee:99,Reithmaier:04,Yoshie:04,Hennessy:07,Gunter:09,Yamanoto:00} to Bose-Einstein condensation of exciton-polaritons~\cite{Deng:02,Kasprzak:06,Balili:07,Deng:10}---have been explored by studying the optical properties of semiconductor microcavities. Yet, the excitation dynamics of the optically induced exciton-polariton states is still largely unexplored. For a GaAs-based microcavity, the Rabi splitting and the exciton binding energy both fall in the range of 1-10 meV corresponding to photon energies in the terahertz (THz) part of the electromagnetic spectrum (i.e., 4.14 meV at 1 THz). Hence, one needs high precision THz spectroscopy to directly access these low-energy processes. Strong and short THz pulses allow for the coherent manipulation of the exciton-polariton states, providing a unique perspective of the coupled light-matter system which is inaccessible in purely optical spectroscopy.

\begin{figure}[h]
\scalebox{0.5}{
\includegraphics{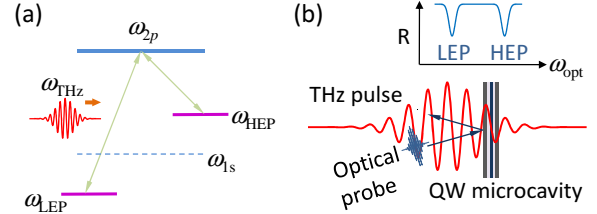}}
\caption{\label{setup} (Color online)
(a) LEP, HEP and exciton 2$p$ polarization states form a three-level $\Lambda$ system.
(b) Schematic diagram of the optical reflectivity measurement of a QW microcavity in the presence of an intense THz field.}
\end{figure}

In this letter, we investigate the quantum dynamics of exciton-polariton coherences in a QW microcavity driven by strong few-cycle THz pulses. We generalize the study of intra-excitonic transitions in bare QWs~\cite{Cole:01,Carter:05,Leinss:08,Danielson:07,Jameson:09} to THz spectroscopy in microcavities, including polaritonic effects. Our experimental observations and theoretical analysis show that a resonant THz field couples the optically induced exciton-polariton coherences to the dark $2p$-exciton state while direct dipole-transitions between the two polariton modes are forbidden. These results indicate that the lower and higher exciton-polariton (LEP and HEP), together with the $2p$-exciton-polarization states constitute a three-level $\Lambda$ system, depicted in Fig.~\ref{setup}(a). This polaritonic $\Lambda$ system is unique and significantly different from the usual atomic $\Lambda$ system where all levels are matter states while all transitions are provided by light. In our case, only the highest $2p$ state is a pure matter state wehreas the LEP and HEP are mixed light-matter coherences. This scenario offers new coherent control possibilities because one can switch the polaritonic LEP and HEP states with optical fields while applying THz fields to perform simultaneous $\Lambda$ transitions. Our results indicate that the novel $\Lambda$ system is robust enough to perform stimulated Raman adiabatic passage (STIRAP)~\cite{Bergmann:98} in a semiconductor microcavity by using two-color nonlinear THz spectroscopy.

As schematically shown in Fig.~\ref{setup}(b), we experimentally observe the time-resolved optical reflectivity of the LEP and HEP modes in a QW microcavity in the presence of strong few-cycle THz pulses.
We combine strong THz and weak optical excitation using 800-nm, 90-fs pulses from a 1-kHz Ti:sapphire regenerative amplifier. The strong few-cycle THz pulses were produced by type-II difference-frequency generation in a 1-mm ZnTe crystal using two linearly-chirped and orthogonally-polarized optical pulses~\cite{Jameson:09,Danielson:08}. The THz pulse duration was 4~ps, and the maximum electric-field amplitude reached 5~kV/cm. The frequency of the high-field THz pulses was continuously tunable from 1.0 to 2.5~THz with a bandwidth of 0.2--0.3~THz. We focused the THz pulses on a QW microcavity sample and measured its optical reflectivity spectra $R(\nu)$ with weak 830-nm, 100-fs optical probe pulses while varying the relative time delay ($\Delta t$) between the THz and optical pulses. The time delay is defined as $\Delta t=t_{\mathrm{THz}}-t_{\mathrm{opt}}$, where a positive time delay means that the THz pulse arrives at the QW later than the optical pulse. The probe pulses were generated with a white-light continuum source and a bandpass filter with 830-nm central wavelength and 10-nm bandwidth. The QW microcavity sample (NMC66) consists of 10 InGaAs QWs in a 11$\lambda$/2-microcavity with distributed Bragg reflectors (DBRs) designed for 99.94\% reflectivity. The temperature of the sample was 5~K.

The experimental results are compared with the theoretical analysis performed with a microscopic many-body theory~\cite{Kira:06, Steiner:08}, including the self-consistent treatment of optical and THz fields. Whereas the full theoretical framework is laid out in the supplementary material~\cite{supp}, we can explained most for the experimental features by including the self-consistent coupling between Maxwell's wave equation and the microscopic interband polarization dynamics

\begin{align} \nonumber
 i \hbar \frac {\partial}{\partial t} P_{\mathbf{k}}
=&  \left(  \tilde \epsilon_{\mathbf{k}}  - j_{ {\mathbf{k}} }  A_{\rm{THz}}
+ \frac {|e|^2}{2 \mu}  A_{\rm{THz}} ^2 \right) P_{\mathbf{k}}
\\ &+ \sum_\mathbf{k'} \Gamma_{\mathbf{k,k'}} P_{\mathbf{k'}}
- \left( 1 - f_{\mathbf{k}} ^e - f_{\mathbf{k}} ^h \right) \Omega_\mathbf{k}.
\end{align}
Here, $\tilde \epsilon_{\mathbf{k}}$ is the kinetic energy of electron-hole pairs and $f^{e(h)}_{\bf k}$ defines the occupation of electrons (holes) at the carrier momentum ${\bf k}$. The optical field generates $P_{\bf k}$ through the renormalized Rabi frequency $\Omega_{\bf k}$ while the microscopic Coulomb- and phonon-interaction induced scattering $\Gamma_{{\bf k},{\bf k}’}$ destroys these coherences. The LEP and the HEP resonances are mixed light-matter states created by the self-consistent coupling between the optically generated $P_{\bf k}$ and the cavity mode which is fully described by the wave equation. The coherent polarization is further coupled to the vector potential $A_{\rm THz}$ of the THz field. The appearing current-matrix element $j({\bf k})$ has a $p$-like symmetry coupling the spherically symmetric optical $P_{\bf k}$ to optically dark exciton states. Interestingly, since the cavity mode is not directly involved in this process, the THz radiation only couples the matter component of the LEP and HEP to the $2p$-exciton state.


\begin{figure}[h]
\scalebox{0.38}{
\includegraphics{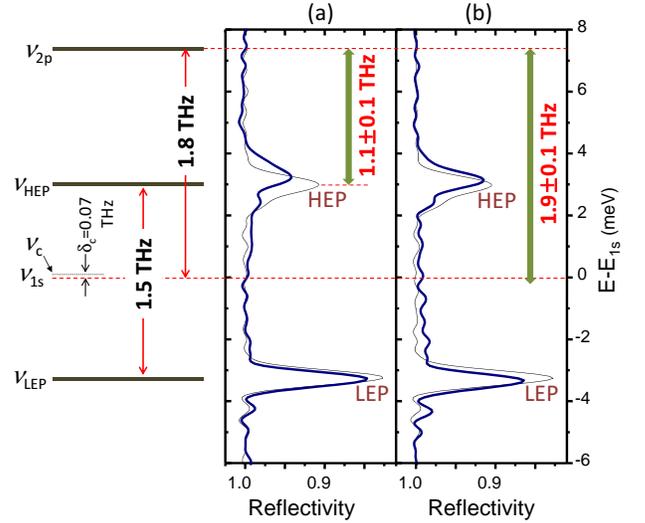}}
\caption{\label{fig2} (Color online)
Exciton-polariton reflectivity spectra at near-zero cavity detuning ($\delta_c=\nu_c-\nu_{1s}=0.07$~THz) in the presence of a strong THz pulse, $E_{\mathrm{THz}}$=4~kV/cm, at $\Delta t$=0.0~ps (thick solid lines): (a) THz frequency ($\nu_{\mathrm{THz}}$=1.1~THz) is tuned to the HEP-to-$2p$ transition ($\nu_{\mathrm{HEP}-2p}$=1.1 THz) and (b) THz frequency ($\nu_{\mathrm{THz}}$=1.9 THz) is tuned to the $1s$-to-$2p$ transition ($\nu_{1s-2p}$=1.8 THz). The thin-solid lines indicate the unperturbed exciton-polariton spectra.}
\end{figure}

Figure~\ref{fig2} shows examples of the THz-induced nonlinear optical effects of the exciton-polariton modes at near-zero cavity-detuning ($\delta_c=\nu_c-\nu_{1s}=0.07$~THz) and at zero time-delay ($\Delta t$=0.0~ps) between the optical and THz pulses. The THz frequency is tuned to (a) the HEP-to-$2p$ transition ($\nu_{\mathrm{HEP}-2p}$=1.1~THz) and (b) the $1s$-to-$2p$ transition ($\nu_{1s-2p}$=1.8~THz).
The peak THz-field amplitudes are 4~kV/cm. The unperturbed reflectivity spectra are depicted by the thin grey lines.

The data exhibits several pronounced nonlinear effects showing that the THz-microcavity interaction is resonant with the HEP-to-2$p$ transition because the $\nu_{{\rm HEP}-2p}$ excitation quenches dominantly the HEP resonance [Fig.~\ref{fig2}(a)] while $\nu_{1s-2p}$ creates only moderate changes [Fig.~\ref{fig2}(b)] to the measured reflectivity. Furthermore, no resonant transition between the LEP and HEP modes ($E_{\mathrm{HEP}}-E_{\mathrm{LEP}}=6.2$~meV=1.5~THz) was observed as the THz frequency was tuned in the range of 1.4--1.6~THz. When the THz frequency was increased to $\nu_{{\rm LEP}-2p}=2.4$~THz, the reflectivity was changed significantly only at the LEP. These observations clearly show that the LEP, HEP, and $2p$-levels form a three-level $\Lambda$ system.

Further evidence for the proposed $\Lambda$ configuration is provided by our $\nu_{1s-2p}$ excitation studies. As we see in Fig.~\ref{fig2}(b) the energetic difference between the LEP and HEP resonances incrases in this case, clearly indicating that the $\Lambda$ system experiences the expected AC Stark shift.

\begin{figure}[h]
\scalebox{0.59}{
\includegraphics{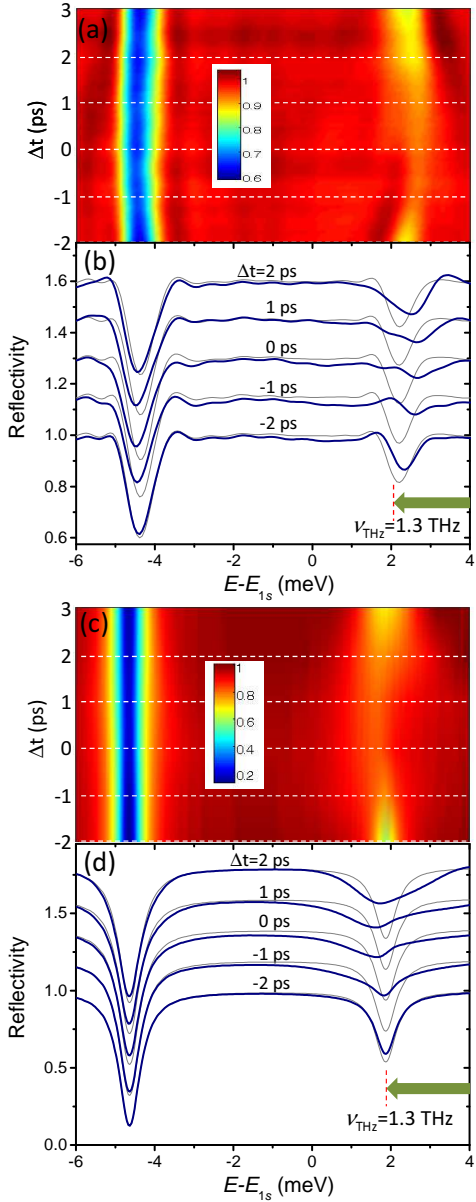}}
\caption{
Time-resolved exciton-polariton reflectivity spectra at $\delta_c$=-0.56~THz when the THz radiation ($\nu_{\mathrm{THz}}$=1.3~THz) is tuned to the HEP-to-$2p$ transition ($\nu_{\mathrm{HEP}-2p}$=1.3~THz).
(a) Experimental and (c) theoretical contour plots of the reflectivity presented with respect to $1s$-exciton energy $E_{1s}$. The white dashed lines correspond to cross-sections at $\Delta t$ = -2, -1, 0, 1, and 2.0~ps shown as vertically offset spectra in (b) experiment and (d) theory. The thin lines indicate the unperturbed spectra.
\label{d-23}}
\end{figure}

\begin{figure}[h]
\scalebox{0.59}{
\includegraphics{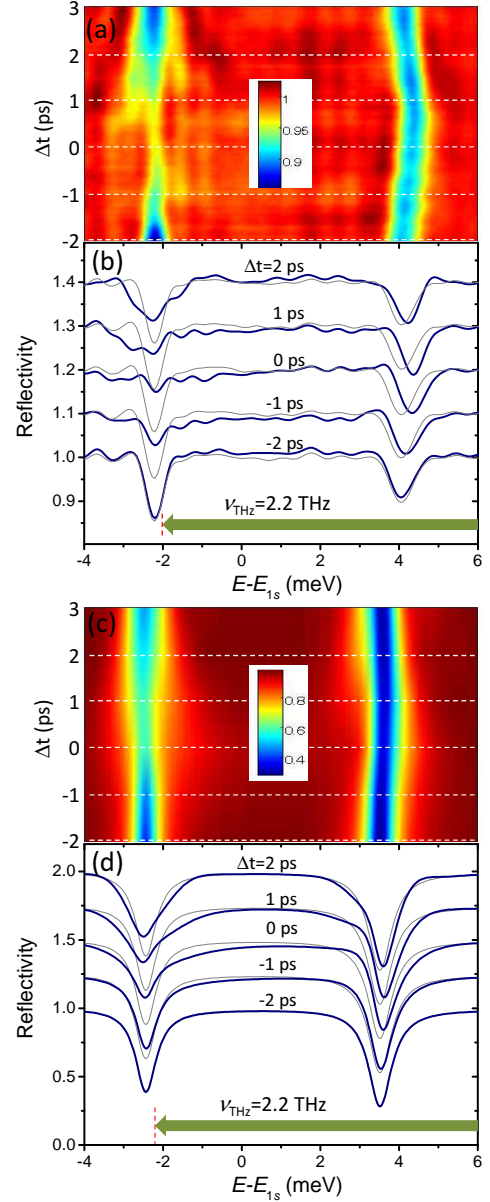}}
\caption{Time-resolved exciton-polariton reflectivity spectra at $\delta_c$=+0.41~THz when the THz radiation ($\nu_{\mathrm{THz}}$=2.2~THz) is tuned to the LEP-to-$2p$ transition ($\nu_{\mathrm{LEP}-2p}$=2.4~THz).
(a) Experimental and (c) theoretical contour plots of the reflectivity $R(E-E_{1s},\Delta t)$. The white dashed lines correspond to cross-sections at $\Delta t$ = -2, -1, 0, 1, and 2.0~ps shown as vertically offset spectra in (b) experiment and (d) theory. The thin lines indicate the unperturbed spectra. \label{d+17}}
\end{figure}

To gain more insights, we performed detailed experiment-theory comparisons for the resonant interactions of THz pulses with the exciton-polariton coherences at different cavity detunings. Figures~\ref{d-23} and \ref{d+17} show the temporal evolution of the exciton-polariton reflectivity when the THz excitation is tuned to the LEP-to-$2p$ transition at $\delta_c=-0.56$~THz and to the HEP-to-$2p$ transition at $\delta_c=0.41$~THz, respectively. Note that the experimental LEP and HEP peaks are lower than the theoretical ones due to the well-known disorder effects in the sample~\cite{Gurioli:01}, yet their bandwidths remain same.

Figures~\ref{d-23}(a) and (c) show the contour plots of the experimental and theoretical reflectivity, $R(E, \Delta t)$, as a function of photon energy $E$ and time delay $\Delta t$ between optical and THz pump, at a negative detuning  $\delta_c$=-0.56~THz. The corresponding reflectivity spectra at $\Delta t$=-2, -1, 0, 1, and 2~ps are shown in Figures~\ref{d-23}(b) and (d). The results indicate that the THz pulses resonantly drive the HEP-to-$2p$ transition, i.e. in agreement with the expectatins for the three-level $\Lambda$ system, the HEP mode undergoes large amplitude modulations near $\Delta t$=0 while little changes occur in the LEP mode. Especially, the HEP resonance almost vanishes near $\Delta t$ = 0 while the LEP remains almost unchanged. This level of quantum control can be observed up to about $\Delta t=2$\,ps, i.e. in the coherent regime before the dephasing of the polariton coherences.

The coherent control of the LEP branch of the $\Lambda$ system is studied in Fig.~\ref{d+17} both experimentally [frames (a) and (b)] and theoretically [frames (c) and (d)] by using $\nu_{\mathrm{THz}}=2.2$\,THz and a cavity detuning $\delta_c$=+0.41 THz. As expected for a $\Lambda$ system response, in this case the THz field predominantly bleaches the LEP. It is also notable that the temporal evolution is not symmetric for the HEP (Fig.~\ref{d-23}) and the LEP branches (Fig. 4). For example, the strongest THz-induced modulation occurs between the time delays of 0.0 and 0.5 ps, and the spectral modulation at $\Delta t =2$\,ps is stronger than that at $\Delta t =-2$\,ps. The temporal asymmetry results mainly from the coherent transients: at positive time delays up to several picoseconds, the THz pulse perturbs the coherent polariton modes inducing the spectral modulations to the probe reflectivity, while they make no impact on the polariton modes at negative time delays.

In conclusion, our study shows that the THz radiation resonantly drives the exciton-polariton polarizations giving rise to LEP-to-$2p$ or HEP-to-$2p$ transitions. The exceptionally large nonlinear optical effects induced by the THz pulses exhibit peculiar spectral features unique for the light-matter coupled system. Our experiment-theory comparison also confirms that there are no resonant transitions between LEP and HEP levels. Hence, we demonstrate that the LEP, HEP, and $2p$-exciton states form a unique three-level $\Lambda$ system in an optically excited QW microcavity. The results also indicate the coherent coupling between the exciton-polariton and the exciton $2p$ polarizations that dephase within a few picosecond. This operational window should be long enough to realize $\Lambda$-system applications, such as THz-STIRAP, in a semiconductor microcavity.

Acknowledgements: The OSU work is supported by the National Science Foundation (DMR-1063632) and the Oregon Nanoscience and Microtechnologies Institute. The Marburg work is supported by the Deutsche Forschungsgemeinschaft. The UofA group acknowledges support from the National Science Foundation under AMOP (PHY-0757707), EPDT (ECCS-0757975) and ERC CIAN (EEC-0812072); AFOSR (FA9550-10-1-0003); and JSOP (W911NF-10-1-0344).



\end{document}